\documentclass[a4paper,12pt]{article}
\usepackage{amsmath,amsfonts,amssymb,cite}
\usepackage[dvips]{graphicx}

\begin{document}

\begin{center}
{\LARGE\bf The sigma meson from QCD sum rules for large-$N_c$
Regge spectra}
\end{center}
\bigskip
\begin{center}
{\large S. S. Afonin
and T. D. Solomko}
\end{center}

\begin{center}
{\it Saint Petersburg State University, 7/9 Universitetskaya nab.,
St.Petersburg, 199034, Russia}
\end{center}
\bigskip
\begin{abstract}
The QCD sum rules in the large-$N_c$ limit for the light
non-strange vector, axial and scalar mesons are considered
assuming a string-like linear spectrum for the radially excited
states. We propose a improved method for combined analysis of
these channels that gives a reasonable description of the observed
spectrum. In the vector-axial case, fixing the pion decay constant
and the gluon condensate we obtain more or less physical values
for the masses of ground states and the quark condensate. Thus a
typical for this method need to fix the mass of some ground state
is overcome. Using in the scalar channel the values of presumably
universal slope of radial trajectories and the quark condensate
obtained in the vector-axial channel, we find that, in contrast to
some strong claims in the literature, a prediction of light scalar
state with a mass close to the mass of $f_0(500)$ can take place
within the method of planar QCD sum rules and may follow in a
natural way from the Regge phenomenology.
\end{abstract}

\section{Introduction}

It is widely known that the physics of non-perturbative strong
interactions is encoded in the hadron masses. This largely
unknown physics is most pronounced in the hadrons consisting of
$u$- and $d$-quarks as the masses $m_{u,d}$ are much less than
the non-perturbative scale $\Lambda_{\text{QCD}}$. At the same time,
these hadrons shape the surrounding world. Aside from the
nucleons and pions, an important role is played by the scalar
$\sigma$-meson which is responsible for the main part of the
nucleon attraction potential. In the particle physics, the given
resonance is identified as $f_0(500)$~\cite{pdg} and is indispensable
for description of the chiral symmetry breaking in many
phenomenological field models for the strong interactions.
In spite of the great efforts invested in the study of this
non-ordinary resonance in the last 60 years, its nature remains
disputable~\cite{pelaez}.

The physical characteristics of hadrons are encoded in various
correlation functions of corresponding hadron currents. Perhaps
the most important characteristics is the hadron mass. The
calculation of a hadron mass from first principles consists in
finding the relevant pole of two-point correlator $\left\langle
JJ\right\rangle$, where the current $J$ is built from the quark
and gluon fields and interpolates the given hadron. For instance,
if the scalar isoscalar state $f_0$ represents an ordinary light
non-strange quark-antiquark meson, its current should be
interpolated by the quark bilinear $J=\bar{q}q$, where $q$ stays
for the $u$ or $d$ quark. In the real QCD, the straightforward
calculations of correlators are possible only in the framework of
lattice simulations which are still rather restricted.

A well-known phenomenological way for extraction of masses and
other characteristics from the correlators is provided by various
QCD sum rules. This method exploits some information from QCD via
the Operator Product Expansion (OPE) of correlation
functions~\cite{svz}. On the other hand, one assumes a certain
spectral representation for a correlator in question. Typically
the representation is given by the ansatz "one infinitely narrow
resonance + perturbative continuum". Such an approximation is very
rough but works well phenomenologically in many
cases~\cite{svz,rry,snapshot,colangelo}. From the theoretical
viewpoint, the zero-width approximation (and simultaneously the
absence of multiparticle cuts) arises in the large-$N_c$ (called
also planar) limit of QCD~\cite{hoof,witten}. In this limit, the
only singularities of the two-point correlation function of a
hadron current $J$ are one-hadron states~\cite{witten}. In the
case of mesons, the two-point correlator has the following form to
lowest order in $1/N_c$ (in the momentum space),
\begin{equation}
\label{1}
\left\langle J(q)J(-q)\right\rangle=\sum_n\frac{F_n^2}{q^2-M_n^2},
\end{equation}
where the large-$N_c$ scaling of quantities is:
$M_n=\mathcal{O}(1)$ for masses,
$F_n^2=\langle0|J|n\rangle^2=\mathcal{O}(N_c)$ for residues,
$\Gamma=\mathcal{O}(1/N_c)$ for the full decay width~\cite{witten}. Due to
asymptotic freedom, the left-hand side of~\eqref{1} behaves
logarithmically at large $q^2$. This behavior is only possible if
the number of terms in the sum is infinite~\cite{witten}.

The logarithmic behavior of the right-hand side of~\eqref{1}
emerges naturally if one has the following large-$n$ asymptotics:
$F^2_n\sim\text{const}$, $M_n^2\sim \Lambda^2 n$. Such a
Regge-like behavior for masses of radially excited states appears
in the two-dimensional QCD in the planar limit~\cite{hoof2},
Veneziano dual amplitudes~\cite{avw}, and various hadron string
models~\cite{string}. In addition, the relation
$F^2_n=\text{const}$ can be regarded as a natural consequence of
the string picture even without assumption on Regge
behavior~\cite{pmc}. Within the aforementioned approaches, the
slope $\Lambda^2$ is independent of the quantum numbers. This can
be explained by universality of gluodynamics which determines the
slope. The radial Regge behavior in the light non-strange mesons
has some experimental evidence~\cite{phen,phen2}. The experimental
slopes do demonstrate an approximately universal behavior. Within
the accuracy of the large-$N_c$ limit (10 - 20\%), the
universality of slopes is a quite adequate assumption.

Considering the linear ansatz $M_n^2 = \Lambda^2 n + M_0^2$ for
the radial mass spectrum, the sum in~\eqref{1} can be summed up,
expanded at large $Q^2=-q^2$ and compared with the corresponding
OPE in QCD. The ensuing planar sum rules were considered many
times in the past (see, e.g.,~\cite{sr,peris,we,AE,arriola}).
Later it became clear that the given sum rules are tightly related
with a popular bottom-up holographic approach to QCD (see, e.g.,
discussions in~\cite{holog2010}). On the other hand, the
phenomenological understanding of spectral regularities has
improved recently (an incomplete list of references
is~\cite{phen3,phen4,phen5}). It seems timely to refresh the
method of planar sum rules and exploit it again in the hadron
phenomenology.

The main focus of our work will be concentrated on the enigmatic
$\sigma$-meson. It is usually believed that the mass of the
lightest scalar quark-antiquark state lies near 1~GeV or
higher~\cite{rry,pelaez}. The $\sigma$-meson, also referred to as $f_0(500)$ in
Particle Data~\cite{pdg}, is much lighter. Various phenomenological
approaches insist on a highly unusual (likely tetraquark) nature
of $\sigma$-particle~\cite{pelaez}. Our intention was to confirm
the absence of a light scalar particle among usual mesons using
the QCD sum rules in the large-$N_c$ limit combined with the Regge
phenomenology. Our conclusion, however, turned out to be opposite
--- a light ordinary scalar state can be predicted in a
natural way within the considered framework. We will also comment
briefly why this result was not obtained earlier in various QCD
sum rules.

The paper is organized as follows. In Section~2, we recall the
derivation of planar sum rules in the vector case. This derivation
is extended to the axial channel in Section~3. In Section~4,
we propose a solution of combined vector-axial sum rules. This
solution is then used in the scalar channel in Section~5.
Section~6 is devoted to some discussions. We conclude in Section~7.

\section{Vector mesons}

Due to conservation of the vector current
$J_\mu^V=\bar{q}\gamma_\mu q$, the vector two-point correlator is
transverse and depends on one scalar function only,
\begin{equation}
\label{2}
\left\langle J_\mu^V(q)J_\nu^V(-q)\right\rangle=(q_\mu q_\nu-g_{\mu\nu}q^2)\Pi_V(q^2).
\end{equation}
Following the discussions in Introduction, we will assume the
simplest linear Regge ansatz for the vector spectrum,
\begin{equation}
\label{3}
M_V^2(n)=\Lambda^2n+M_V^2, \qquad n=0,1,2,\dots.
\end{equation}
Since the isosinglet and isotriplet states are degenerate in
the large-$N_c$ (and chiral) limit~\cite{witten}, the spectra of $\omega$
and  $\rho$ mesons are indistinguishable in our framework.
We will discuss the isosinglet states.

There are at least two reasons to separate the ground state out of
the linear trajectory~\eqref{3}. First, the available experimental
data show that the ground state lies noticeably below the linear
trajectory in all unflavored vector quarkonia~\cite{phen5}. An
example for the $\omega$-mesons is depicted in Fig.~1. Second, the
ground $\omega$ and $\rho$ mesons belong to the leading angular
Regge trajectory. It is known that the meson states on this
trajectory do not have parity (and chiral) partners~\cite{phen3}.
Hence, the vector channel should have one additional state with
respect to the axial channel which will be considered in the next
Section.

Using the spectral representation~\eqref{1}, definition~\eqref{2},
and ansatz~\eqref{3} we get in the euclidean domain $Q^2=-q^2$,
\begin{equation}
\label{4}
\Pi_V(Q^2)=\frac{F_\omega^2}{Q^2+M_\omega^2}+\sum_{n=0}^{\infty}\frac{F^2}{Q^2+\Lambda^2n+M_V^2}.
\end{equation}
As we motivated in Introduction, the residues of excited states
in~\eqref{4} are assumed to be constant and universal. In addition,
it can be easily demonstrated that the asymptotics "logarithm
+ power terms"~\eqref{5} holds only if $F^2_n\sim\frac{dM^2_n}{dn}$~\cite{we}
which gives a constant for the linear ansatz~\eqref{3}.

In the chiral and planar limits (with setting $N_c=3$ at the end),
the Operator Product Expansion (OPE) of the vector correlator at
large $Q^2$ reads~\cite{svz}
\begin{equation}
\label{5}
\Pi_V(Q^2)=-\frac{C_0}{8\pi^2}\log{\frac{Q^2}{\mu^2}}+
\frac{\alpha_s}{24\pi}\frac{\langle G^2\rangle}{Q^4}
-\frac{14}{9}\pi\alpha_s\frac{\langle\bar{q}q\rangle^2}{Q^6}+\dots,
\end{equation}
where $\langle G^2\rangle$ and $\langle\bar{q}q\rangle$ denote the
gluon and quark vacuum condensate, respectively. According to the
tenets of classical QCD sum rules~\cite{svz}, these vacuum
characteristics are universal, i.e., their values do not depend on
the quantum numbers of a hadron current $J$ (the method is not
applicable otherwise). The factor $C_0$ includes the perturbative
correction to the leading logarithm, $C_0=1+\frac{\alpha_s}{\pi}$.
Within the accuracy of the large-$N_c$ limit, the correction is
rather small and cannot be taken into account reliably. We set
$C_0=1$ in what follows.

The expression~\eqref{4} can be rewritten via the $\psi$-function
(a logarithmic derivative of $\Gamma$-function),
\begin{equation}
\label{6}
\sum_{n=0}^{\infty}\frac{1}{n+a}=-\psi(a)+\text{const},
\end{equation}
which has the following asymptotic expansion at large argument,
\begin{equation}
\label{7}
\psi(z)=\log{z}-\frac{1}{2z}-\sum_{k=1}^{\infty}\frac{B_{2k}}{2kz^{2k}}.
\end{equation}
Here $B_{2k}$ are Bernulli numbers. With the help of these
formulas, the correlator~\eqref{4} can be expanded
at large $Q^2$. In terms of the dimensionless variables
\begin{equation}
\label{8}
m_v=\frac{M_V}{\Lambda}, \qquad m_\omega=\frac{M_\omega}{\Lambda}, \qquad
f=\frac{F}{\Lambda}, \qquad f_\omega=\frac{F_\omega}{\Lambda},
\end{equation}
the result is
\begin{multline}
\label{9}
\Pi_V(Q^2)=-f^2\log{\frac{Q^2}{\mu^2}}+
\frac{\Lambda^2}{Q^2}\left[f_\omega^2-f^2\left(m_v^2-\frac12\right)\right]+\\
\frac{\Lambda^4}{Q^4}\left[-f_\omega^2m_\omega^2+\frac12f^2\left(m_v^4-m_v^2+\frac16\right)\right]+\\
\frac{\Lambda^6}{Q^6}\left[f_\omega^2m_\omega^4-\frac13f^2m_v^2\left(m_v^2-\frac12\right)\left(m_v^2-1\right)\right]+\dots.
\end{multline}

The planar sum rules for the linear spectrum~\eqref{3} follow from
the comparison of~\eqref{9} with~\eqref{5}. But first let us
consider the axial-vector channel.

\section{Axial mesons}

As the axial-vector current $J_\mu^A=\bar{q}\gamma_\mu\gamma_5 q$
is not conserved, the axial two-point correlator has two
independent contributions,
\begin{equation}
\label{10}
\left\langle J_\mu^A(q)J_\nu^A(-q)\right\rangle=\Pi_A(q^2)q_\mu q_\nu-\tilde{\Pi}_A(q^2)g_{\mu\nu}.
\end{equation}
The sum rules for $\Pi_A$ and $\tilde{\Pi}_A$ are different
because the longitudinal part $\Pi_A$ contains an extra
contribution from the pion pole due to PCAC, $J^A_\mu\sim
f_\pi\partial_\mu\pi$. In our normalization, the value of the pion
weak decay constant is $f_\pi=93$~MeV. Since the classical
Weinberg paper~\cite{wein} one traditionally extracts the transverse
part in~\eqref{10} (by adding and subtracting the term
$g_{\mu\nu}q^2\Pi_A$) and considers the sum rules for $\Pi_A$ in
conjunction with the sum rules for $\Pi_V$.

As was motivated in Introduction, we assume a linear ansatz for
the radial axial spectrum with universal slope. The axial analogue
of the correlator~\eqref{4} is
\begin{equation}
\label{11}
\Pi_A(Q^2)=\frac{f_\pi^2}{Q^2}+\sum_{n=0}^{\infty}\frac{F^2}{Q^2+\Lambda^2n+M_A^2}.
\end{equation}
Strictly speaking, we should consider the isosinglet $\eta$-meson
in place of the pion. In the two-flavor case,
however, the difference is not substantial.
The OPE of the correlator~\eqref{11} reads~\cite{svz}
\begin{equation}
\label{12}
\Pi_A(Q^2)=-\frac{C_0}{8\pi^2}\log{\frac{Q^2}{\mu^2}}+
\frac{\alpha_s}{24\pi}\frac{\langle G^2\rangle}{Q^4}
+\frac{22}{9}\pi\alpha_s\frac{\langle\bar{q}q\rangle^2}{Q^6}+\dots.
\end{equation}
It should be noted that only the last term in~\eqref{5}
and~\eqref{12} is different. Proceeding further as in the vector
case, in terms of dimensionless notations~\eqref{8}
($m_a=\frac{M_A}{\Lambda}$) we get
\begin{multline}
\label{13}
\Pi_A(Q^2)=-f^2\log{\frac{Q^2}{\mu^2}}+
\frac{\Lambda^2}{Q^2}\left[\frac{f_\pi^2}{\Lambda^2}-f^2\left(m_a^2-\frac12\right)\right]+\\
\frac{\Lambda^4}{Q^4}\frac{f^2}{2}\left(m_a^4-m_a^2+\frac16\right)-
\frac{\Lambda^6}{Q^6}\frac{f^2}{3}m_a^2\left(m_a^2-\frac12\right)\left(m_a^2-1\right)+\dots.
\end{multline}

As in the vector case, the pure axial sum rules follow from
comparison of~\eqref{12} with~\eqref{13}.

\section{Vector sum rules}

As was indicated above, the combined set of vector-axial sum rules
emerges from equating terms at $\log{Q^2}$, $1/Q^2$, $1/Q^4$, and
$1/Q^6$ in~\eqref{5} and~\eqref{9} and in~\eqref{12}
and~\eqref{13}. Our inputs will be the pion decay constant $f_\pi$
and the gluon condensate $\frac{\alpha_s}{\pi}\langle G^2\rangle$.
The quark condensate will be a prediction. More precisely, we
predict the value of dim-6 condensate
$\alpha_s\langle\bar{q}q\rangle^2$ which has a rather small but
non-zero anomalous dimension. The sum rules are consistent at some
definite value of the dim-6 condensate. The quark condensate at
certain normalization point can be deduced from this value. Thus
at $1/Q^6$ we will have only one sum rule which follows from
equating the $1/Q^6$-terms in~\eqref{9} and~\eqref{13} with the
factor $-7/11$ (as prescribed by the OPE~\eqref{5}
and~\eqref{12}). The resulting set of equations is
\begin{align}
f^2&=\frac{1}{8\pi^2},\\
f^2\left(m_v^2-\frac12\right)&=f_\omega^2,\\
\Lambda^2f^2\left(m_a^2-\frac12\right)&=f_\pi^2,\\
\Lambda^4\left[-f_\omega^2m_\omega^2+\frac12f^2\left(m_v^4-m_v^2+\frac16\right)\right]&=\frac{\alpha_s}{24\pi}\langle G^2\rangle,\\
\Lambda^4f^2\left(m_a^4-m_a^2+\frac16\right)&=\frac{\alpha_s}{12\pi}\langle G^2\rangle,\\
f_\omega^2m_\omega^4-\frac13f^2m_v^2\left(m_v^2-\frac12\right)\left(m_v^2-1\right)&=
\frac{7}{33}f^2m_a^2\left(m_a^2-\frac12\right)\left(m_a^2-1\right).
\end{align}

Thus we arrive at the system of 6 polynomial equations with 6
variables $\Lambda^2$, $m_v^2$, $m_\omega^2$, $m_a^2$, $f^2$, and
$f_\omega^2$. This system can be solved numerically. The values of
inputs are $f_\pi=93$~MeV and $\frac{\alpha_s}{\pi}\langle
G^2\rangle=(360\pm20\,\text{MeV})^4$. To demonstrate the
sensitivity of solutions to a choice of inputs we try also
$f_\pi=87$~MeV (a presumable value of $f_\pi$ in the chiral
limit~\cite{gasser}) and show the uncertainty caused by the
uncertainty in the value of the gluon condensate. The physical
solutions after rescaling~\eqref{8} are given in Table~1.
Concerning the dim-6 condensate, a self-consistent interpretation
of the value of $\langle\bar{q}q\rangle$ appears at the choice
$\alpha_s\simeq1/\pi\simeq0.3$ that corresponds to the scale
$\mu\simeq2$~GeV. So the obtained value of the quark condensate
refers to that scale. The predicted masses of first 3 states are
displayed in Table~2.
\begin{table}
\caption{\small The numerical solutions in GeV (see text).}
\begin{center}
{
\begin{tabular}{|c|cc|}
\hline
 & $f_\pi=93$~MeV  & $f_\pi=87$~MeV \\
\hline
$\Lambda$ & 1.43(2) & 1.32(2)\\
$M_V$ & 1.60(4) & 1.45(4) \\
$M_A$ & 1.31(1) & 1.21(1)\\
$M_\omega$ & 0.79(3) & 0.69(3)\\
$F$ & 0.16 & 0.15\\
$F_\omega$ & 0.14 & 0.13\\
$(-\langle\bar{q}q\rangle)^{\frac13}$ & 0.30(1) & 0.27(1)\\
\hline
\end{tabular}}
\end{center}
\end{table}

Taking into account all rough approximations that we have made,
the resulting solution is surprisingly good. First of all, the
masses of ground states are close to the experimental masses
of unflavored vector $\omega(782)$ and axial $f_1(1285)$
mesons~\cite{pdg}. The agreement looks excellent for the
large-$N_c$ limit. Second, the obtained value of
$\langle\bar{q}q\rangle$ is also unexpectedly reasonable. Indeed,
the numerical solution yields the product
$\alpha_s\langle\bar{q}q\rangle^2$, where both $\alpha_s$ and the
quark condensate $\langle\bar{q}q\rangle$ depend on the
normalization scale $\mu$. Excluding $\mu$ one can draw a
"physical" curve on the
$\left(\alpha_s,\langle\bar{q}q\rangle\right)$ plane. The fact
that our solution approximately belongs to this curve is
non-trivial.
\begin{table}
\caption{\small The masses of first three predicted states in GeV
(central values).}
\begin{center}
{
\begin{tabular}{|c|ccc|}
\hline
$n$ & 0 & 1 & 2  \\
\hline
$f_\pi=93$~MeV & & &\\
$M_V(n)$ & 0.79 & 1.60 & 2.15 \\
$M_A(n)$ & 1.31 & 1.93 & 2.41\\
\hline
$f_\pi=87$~MeV & & &\\
$M_V(n)$ & 0.69 & 1.45 & 1.96 \\
$M_A(n)$ & 1.21 & 1.79 & 2.22\\
\hline
\end{tabular}}
\end{center}
\end{table}

As to the radially excited states, there is still a large
controversy in interpretation of the relevant experimental data
and in determination of real physical masses~\cite{pdg}. In view
of a rather qualitative character of our model we do not want to
delve into the corresponding speculations. Our predictions refer
to the large-$N_c$ limit. The real masses of excited states must
be shifted by various effects which are beyond the scope of our
model. We just mention that the obtained masses seem to lie in the
correct mass ranges and thereby look reasonable.

\section{Scalar sum rules}

Consider the two-point correlator of the scalar isoscalar current
$J^S=\bar{q}q$. Its resonance representation reads (up to two
contact terms)
\begin{equation}
\label{20}
\Pi_S(q^2)=\left\langle J^S(q)J^S(-q)\right\rangle=\sum_n\frac{G_n^2M_S^2(n)}{q^2-M_S^2(n)},
\end{equation}
where the residues stem from the definition
$\langle0|J^S|n\rangle=G_nM_S(n)$. As in the vector cases, we
assume the linear radial spectrum with universal slope
\begin{equation}
\label{21}
M_S^2(n)=\Lambda^2n+M_S^2, \qquad n=0,1,2,\dots.
\end{equation}
And as in the vector channels, within the linear
ansatz~\eqref{21}, the analogues of decay constant must be equal
for consistency with the OPE: $G_n=G$.

\begin{figure}[ht]
\vspace{-0.8cm}
\center{\includegraphics[width=1\linewidth]{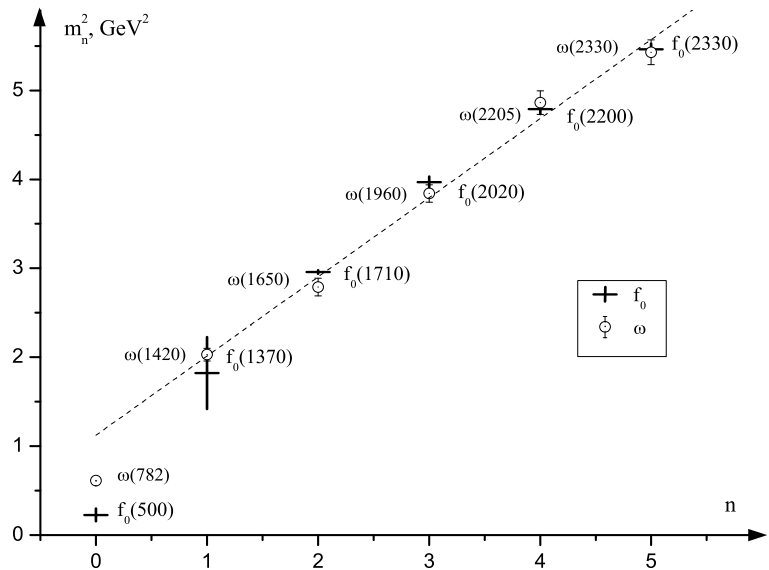}}
\vspace{-1.7cm} \caption{A presumable spectrum of non-strange
$\omega$ (circles) and $f_0$ (crosses) mesons~\cite{pdg}. A rather
large fixed horizontal size of crosses is drawn to indicate better
the position of scalar resonances. The $f_0(1500)$ is excluded as
the available data on this state are poorly compatible with the
$q\bar{q}$-assignment (see the mini-review "Non-$q\bar{q}$ Mesons"
in Particle Data~\cite{pdg}). The plot is taken from
Ref.~\cite{AK}.}
\end{figure}
As {\it apriori} we do not know reliably the radial Regge behavior
of scalar masses, we will consider two simple possibilities: (I)
The ground $n=0$ state lies on the linear trajectory~\eqref{21};
(II) The state $n=0$, below called $\sigma$, is not described by
the linear spectrum~\eqref{21}. The second assumption looks more
physical, see Fig.~1. The corresponding spectral representations
in the Euclidean space are
\begin{equation}
\label{22}
\Pi_S^{(I)}(Q^2)=\sum_{n=0}^{\infty}\frac{G^2(\Lambda^2n+M_S^2)}{Q^2+\Lambda^2n+M_S^2},
\end{equation}
\begin{equation}
\label{23}
\Pi_S^{(II)}(Q^2)=\frac{G_\sigma^2M_\sigma^2}{Q^2+M_\sigma^2}+\sum_{n=1}^{\infty}\frac{G^2(\Lambda^2n+M_S^2)}{Q^2+\Lambda^2n+M_S^2}.
\end{equation}
Proceeding further as in the vector case, we expand~\eqref{22}
and~\eqref{23} at large $Q^2$ and compare the expansions with the
OPE of the scalar correlator~\eqref{20}. Introducing the
dimensionless variables
\begin{equation}
\label{24}
m_s=\frac{M_S}{\Lambda},  \qquad
g=\frac{G}{\Lambda},
\end{equation}
the expansions have the form
\begin{multline}
\label{25}
\Pi_S^{(I)}(Q^2)=g^2Q^2\log{\frac{Q^2}{\mu^2}}-
\frac{\Lambda^4}{Q^2}\frac{g^2}{2}\left(m_s^4-m_s^2+\frac16\right)+\\
\frac{\Lambda^6}{Q^4}\frac{g^2}{3}m_s^2\left(m_s^2-\frac12\right)\left(m_s^2-1\right)+\dots,
\end{multline}
\begin{multline}
\label{26}
\Pi_S^{(II)}(Q^2)=g^2Q^2\log{\frac{Q^2}{\mu^2}}+\frac{G_\sigma^2M_\sigma^2}{Q^2}-
\frac{\Lambda^4}{Q^2}\frac{g^2}{2}\left(m_s^4+m_s^2+\frac16\right)-\\
\frac{G_\sigma^2M_\sigma^4}{Q^4}+
\frac{\Lambda^6}{Q^4}\frac{g^2}{3}m_s^2\left(m_s^2+\frac12\right)\left(m_s^2+1\right)+\dots,
\end{multline}
The OPE of the correlator~\eqref{20} in the chiral and large-$N_c$
limits reads~\cite{rry}
\begin{equation}
\label{27}
\Pi_S(Q^2)=\frac{3C_0}{16\pi^2}Q^2\log{\frac{Q^2}{\mu^2}}+
\frac{\alpha_s}{16\pi}\frac{\langle G^2\rangle}{Q^2}
-\frac{11}{3}\pi\alpha_s\frac{\langle\bar{q}q\rangle^2}{Q^4}+\dots,
\end{equation}
where
\begin{equation}
\label{27b}
C_0=1+\frac{11\alpha_s}{3\pi}.
\end{equation}
Now the perturbative
correction can contribute more than 30\% to the factor in front of
the logarithm. This contribution has a much stronger
impact than in the vector channels and should be taken into account.
Matching the logarithmic terms we obtain
\begin{equation}
\label{28}
g^2=\frac{3C_0}{16\pi^2}.
\end{equation}

Consider the assumption (I). From~\eqref{25} and~\eqref{27} we
have two sum rules,
\begin{align}
\label{29}
\frac{3C_0}{2\pi^2}\Lambda^4\left(m_s^4-m_s^2+\frac16\right)&=-\frac{\alpha_s}{\pi}\langle G^2\rangle,\\
\frac{3C_0}{16\pi^2}\Lambda^6m_s^2\left(m_s^2-\frac12\right)\left(m_s^2-1\right)&=
-11\pi\alpha_s\langle\bar{q}q\rangle^2.
\label{30}
\end{align}
Substituting the numerical values of $\Lambda$ and
$\langle\bar{q}q\rangle$ from the solution of vector sum rules
(Table~1), we arrive at two independent polynomial equations. If
we neglect the perturbative correction in~\eqref{27b}, $C_0=1$, the
equations~\eqref{29} and~\eqref{30} share an approximately common
solution\footnote{At $f_\pi=93$~MeV. Using $f_\pi=87$~MeV, the
solution is $m_s^2\simeq0.72$ resulting in the spectrum
$M_S(n)\simeq1.12, 1.73, 2.17, \dots$~GeV.} $m_s^2\simeq0.74$
leading to the radial scalar spectrum $M_S(n)\simeq1.23, 1.89,
2.37, \dots$~GeV. If we include the perturbative correction, a
miracle with the common solution disappears.

Consider a more physical assumption (II). Matching~\eqref{26} with
the OPE~\eqref{27} leads to the following sum rules,
\begin{align}
\label{31}
G_\sigma^2M_\sigma^2-\frac{3C_0}{32\pi^2}\Lambda^4\left(m_s^4+m_s^2+\frac16\right)&=\frac{\alpha_s}{16\pi}\langle G^2\rangle,\\
-3G_\sigma^2M_\sigma^4+\frac{3C_0}{16\pi^2}\Lambda^6m_s^2\left(m_s^2+\frac12\right)\left(m_s^2+1\right)&=
-11\pi\alpha_s\langle\bar{q}q\rangle^2.
\label{32}
\end{align}
Now we have two equations with three variables $m_s$, $M_\sigma$,
and $G_\sigma$. Excluding $G_\sigma$ we get a relation for the
mass of $\sigma$-meson as a function of the intercept parameter
$m_s^2$,
\begin{equation}
\label{33}
M_\sigma^2=\frac{\frac{C_0}{16\pi^2}\Lambda^6m_s^2\left(m_s^2+\frac12\right)\left(m_s^2+1\right)+
\frac{11}{3}\pi\alpha_s\langle\bar{q}q\rangle^2}
{\frac{3C_0}{32\pi^2}\Lambda^4\left(m_s^4+m_s^2+\frac16\right)+\frac{\alpha_s}{16\pi}\langle G^2\rangle}.
\end{equation}
The "decay constant" $G_\sigma$ as a function of $m_s^2$ can be
obtained by substituting~\eqref{33} to~\eqref{31} or~\eqref{32}.
The quantities $M_\sigma$, $G_\sigma$, $G=\Lambda g$ (where $g$
is defined in~\eqref{28}), and mass of the first state on the scalar
trajectory are plotted in Fig.~2 using the inputs
from Table~1 for $f_\pi=93$~MeV and $\alpha_s\simeq1/\pi$
in~\eqref{27b} that was obtained in the vector case.
The intercept $m_s^2$ can be
negative as the sum in~\eqref{23} begins with $n=1$.
\begin{figure}[ht]
\center{\includegraphics[width=0.7\linewidth]{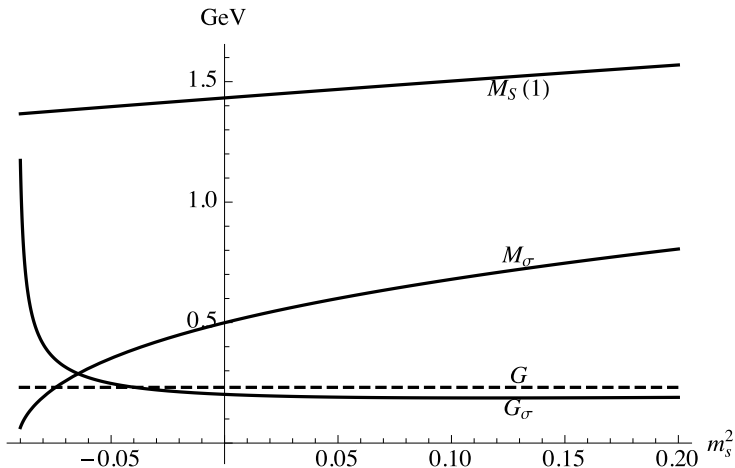}}
\vspace{-0.3cm}
\caption{The values of $M_\sigma$, $G_\sigma$, $G$, and
the first state on the scalar trajectory $M_S(1)$ as a
function of dimensionless intercept $m_s^2$ from the
relations~\eqref{33},~\eqref{31} (or~\eqref{32}),~\eqref{28},
and $M_S^2(1)=\Lambda^2(1+m_s^2)$ from~\eqref{21}.}
\end{figure}

We checked also other variants with inputs corresponding to
$f_\pi=87$~MeV in Table~1 and with $\alpha_s=0$ in~\eqref{27b}.
They result in a shift within 70-80~MeV for masses that lies
within the accuracy of the large-$N_c$ limit. The general picture
displayed in Fig.~2 remains, however, the same for all variants.
Going to negative intercept an unphysical behavior emerges already
at relatively small values. The mass $M_S(1)$ is rather stable and
seems to reproduce the mass of $a_0(1450)$-meson,
$M_{a_0(1450)}=1474\pm19$~MeV~\cite{pdg}. Its isosinglet partner
(the candidates is $f_0(1370)$) should be degenerate with
$a_0(1450)$ in the planar limit.

The plot in Fig.~2 demonstrates that the actual prediction for
$M_\sigma$ is very sensitive to the intercept of scalar linear
trajectory, though by assumption $M_\sigma$ is not described by
the linear spectrum~\eqref{21}. And {\it vice versa}, the expected
value of $M_\sigma$ (around $0.5$~GeV~\cite{pdg}) imposes a strong
bound on the allowed values of intercept $m_s^2$. The plot in
Fig.~2 shows that $m_s^2$ is close to zero.

Although both the ground $\omega$-meson and $\sigma$ lie out of
the corresponding linear trajectories (as is suggested, e.g., by
Fig.~1) there is a difference between them in our analysis. In the
vector case, it was important to start the sum in~\eqref{4} from
$n=0$ in order to relate the resonance representation in the
vector case to the axial one in~\eqref{11}. If we started from
$n=0$ in the scalar channel~\eqref{23}, the sign of both numerator
and denominator in~\eqref{33} would depend on the value of
$m_s^2$,
\begin{equation}
\label{34}
M_\sigma^2=\frac{\frac{C_0}{16\pi^2}\Lambda^6m_s^2\left(m_s^2-\frac12\right)\left(m_s^2-1\right)+
\frac{11}{3}\pi\alpha_s\langle\bar{q}q\rangle^2}
{\frac{3C_0}{32\pi^2}\Lambda^4\left(m_s^4-m_s^2+\frac16\right)+\frac{\alpha_s}{16\pi}\langle G^2\rangle},
\end{equation}
making the prediction of $M_\sigma$ highly unstable and uncertain.
In this sense, the $\sigma$-meson is not unusual since it belongs
to the radial scalar trajectory. Just its mass is not described by
the linear ansatz~\eqref{21}. The given interpretation can be also
motivated by a comparison of residues --- $G_\sigma$ lies only
slightly below $G$. Physically this means that an external source
(some scalar current) of scalar mesons creates the lightest state
with a probability close to the probabilities of creation of other
scalar resonances. Within our accuracy, the coupling of
$\sigma$-meson to that source is barely suppressed.

The observation above suggests to check the hypothesis
$G_\sigma=G$ explicitly. After this substitution, the sum
rules~\eqref{31}--\eqref{32} (or their analogues if we start from
$n=0$ in~\eqref{23}, the shift does not change the results)
represent two equations with two unknown variables $M_\sigma^2$
and $m_s^2$. This system has 4 numerical solutions. Two of them
are unphysical (give tachionic masses). The third one is (in terms
of dimensionless quantities $m_\sigma=M_\sigma/\Lambda$,
$m_s=M_S/\Lambda$ and for $f_\pi=93$~MeV):
$(m_\sigma^2,m_s^2)\simeq(0.742,0.739)$. This case reproduces the
solution of the assumption (I) above. The given solution
corresponds to the branch where the condensates in the r.h.s. of
Eqs.~\eqref{31}--\eqref{32} can be neglected. The fourth solution,
$(m_\sigma^2,m_s^2)\simeq(0.074,-0.040)$, predicts
$M_\sigma\simeq0.39$~GeV and the radial spectrum
$M_S(n)\simeq1.40, 2.01, \dots$~GeV. These values (except masses
of higher excitations) are visually seen in Fig.~2 --- they
correspond to the point where the lines $G$ and $G_\sigma$
intersect. The given solution is the most interesting: The
obtained mass of $\sigma$-meson lies close to the expected mass
range~\cite{pdg} and the radial spectrum looks reasonable. For
instance, the first radial state can be identified with
$f_0(1370)$ in Fig.~1. This state has a natural isovector partner
$a_0(1450)$~\cite{pdg}. Also the second radial excitation has a
natural interpretation --- the resonance $f_0(2020)$ in Fig.~1.
Setting $f_\pi=87$~MeV, the corresponding predictions are:
$M_\sigma\simeq0.38$~GeV and $M_S(n)\simeq1.30, 1.85, \dots$~GeV.

When one predicts some quark-antiquark state it is important to
indicate its place on the angular Regge trajectory as well. In
other words, what are $f_2,\,f_4,\,\dots$ companions of $f_0(500)$
on this trajectory? In order to answer this question we must know
the slope of the trajectory under consideration. According to the
analysis of Ref.~\cite{phen}, the slope of $f_0$ trajectory, most
likely, lies in the interval $1.1\div1.2$ GeV$^2$. Several
independent estimations made in some papers of
Ref.~\cite{phen3,phen4} seem to confirm this value. Consider our
preferable estimate on the $\sigma$-meson mass obtained above,
$m_\sigma\approx390$~MeV. Then we obtain
$m_{f_2}\approx1.53\div1.60$~GeV. The PDG contains a well
established resonance $f_2(1565)$~\cite{pdg} with mass
$m_{f_2(1565)}=1562\pm13$~MeV. It is a natural companion of
$\sigma$-meson on the corresponding angular Regge trajectory. The
next state would have the mass $m_{f_4}\approx2.13\div2.23$~GeV.
The discovery of the predicted tensor meson $f_4$ (and perhaps the
next companion $f_6$ with $m_{f_6}\approx2.60\div2.71$~GeV) would
confirm our conjecture about the form of Regge trajectory with the
$\sigma$-meson on the top. A tentative candidate for our $f_4$ in
the Particle Data is the resonance $f_J(2220)$ having still
undetermined spin --- its value is either $J=2$ or
$J=4$~\cite{pdg}. Our model would favor the second possibility.

It is interesting to note that the predicted trajectory is drawn
in Ref.~\cite{phen} among numerous angular Regge trajectories for
isosinglet $P$-wave states of even spin. But the resonance
$f_2(1565)$ is replaced there by $f_2(1525)$ (and is absent on
other trajectories). As a result, $m_{f_0}^2$ has a very small
negative value leading to disappearance of a scalar state from
this trajectory. The predicted $f_4$-companion is labelled as
$f_4(2150)$~\cite{phen}. The modern PDG contains the state
$f_2'(1525)$ but this resonance is typically produced in reactions
with $K$-mesons that evidently indicates on the dominant strange
component. For this reason we should exclude it from our
estimates.

Our prediction of the Regge trajectory containing the
$\sigma$-meson on the top seems to contradict to studies of the
$\sigma$-state on the complex Regge trajectory which claim that
because of very large width the corresponding state cannot belong
to usual Regge trajectories~\cite{pelaez,londergan}. It is not
excluded, however, that this observation may simply indicate on
limitations of the usual methods which are applied to description
of the $\pi\pi$-scattering. These methods are based on analyticity
and unitarity of $S$-matrix and do not contain serious dynamical
inputs. The generation of a huge width for $f_0(500)$ represents,
most likely, some dynamical effect. For this reason genuine nature
of $\sigma$-meson can be uncovered only within dynamical
approaches.

Thus our analysis demonstrates that the existence of a light
scalar state is well compatible with the structure of the planar
sum rules in the scalar channel and may follow in a natural way
from the Regge phenomenology.

\section{Discussions}

There exists a widespread belief that a natural mass of the
lightest quark-antiquark scalar state in the QCD sum rules lies
near 1~GeV. This prediction follows both from the standard
borelized spectral sum rules~\cite{rry} and from the planar sum
rules~\cite{we,AE}. It should be emphasized that the  given
prediction is not definitive but rather represents a consequence
of some specific assumptions and tricks. As was demonstrated in a
recent paper~\cite{sigma}, if one uses the Borel transform and the
typical ansatz "one narrow resonance plus continuum", the
extracted mass of the quark-antiquark scalar state cannot be less
than about 0.8~GeV independently of any further assumptions. This
turns out to be a specific internal restriction of the method
itself\footnote{Another objection against the $\sigma$-meson
within this method was the observation that unmixed scalar
quarkonia ground states are not wide~\cite{narison}. We note that
this result also refers to a scalar state near 1~GeV where the
exploited method is reliable.}. In the planar sum rules, the
reason was different. In case of Ref.~\cite{we}, the result seems
to be related to the fact that one studied the scalar sum rules in
conjunction with the pseudoscalar ones with some shared
parameters. In the considered scheme, the ground scalar state
cannot be significantly lighter than $\pi(1300)$ whose mass was
taken as an input. The pseudoscalar channel is notoriously
problematic and the applicability of the sum rules in this channel
is questionable~\cite{svz,rry}. Thus the assumption made in
Ref.~\cite{we} was rather strong. In the planar analysis of
Ref.~\cite{AE}, the resonance $f_0(980)$ was placed as the first
state on the scalar trajectory and alternative possibilities were
not studied.

In our consideration, the assumptions above are not used. Making
the standard sum rule analysis of the two-point correlator for the
simplest quark-antiquark scalar current in the planar limit, we
have demonstrated that the existence of scalar state compatible
with $f_0(500)$ can be rather natural. But a concrete prediction
for its mass is uncertain, mainly because the form of experimental
radial scalar trajectory is controversial. We have advocated that
the most consistent value of $m_\sigma$ within our scheme lies
near $m_\sigma\approx0.4$~GeV. One should keep in mind that our
predictions refer to the large-$N_c$ limit where meson mixings and
decays are suppressed. In the real world with $N_c=3$, a strong
coupling to two pions should enhance the observable mass of
$\sigma$-meson. A phenomenological way to exclude the mixing with
other meson (typically pion) states in the propagation of
resonances consists in extracting the $K$-matrix poles where the
corresponding "bare states" emerge. Albeit the procedure is
model-dependent, it could make sense to compare the large-$N_c$
masses with the relevant $K$-matrix poles. For instance, the
relevant scalar radial trajectory in Ref.~\cite{phen} has
$f_0(1300)$ (called $f_0(1370)$ in the PDG~\cite{pdg}) on the top.
The corresponding "bare" trajectory, according to
Ref.~\cite{phen}, has a scalar state with the mass
$m_{f_0(\text{bare})}=1240\pm50$~MeV on the top. The slope of
"bare" trajectory is about $\Lambda^2\approx1.38$~GeV$^2$. We
propose to interpret the $\sigma$-meson as the lightest state on
this trajectory. Extending the "bare" trajectory to lower mass, we
obtain an approximate estimate:
$m_\sigma\approx\sqrt{m^2_{f_0(\text{bare})}-\Lambda^2}\approx400\pm100$~MeV.
This estimate agrees with our result. In Ref.~\cite{phen},
however, the $\sigma$-meson was claimed to be alien to the
classification of $\bar{q}q$-states.

Since the used sum rule method is based on the narrow-width
approximation, a direct translation of our predictions to the
physical parameters of a broad resonance looks questionable. As a
matter of fact, we claim only that a scalar isoscalar pole in the
range 400--600 MeV can naturally exist in the large-$N_c$ limit.

Another pertinent question is why the $\sigma$-meson lies below
the linear radial Regge trajectory like the ground vector states?
In the latter case, one can propose a simple qualitative
explanation. The ground vector states are $S$-wave, so they
represent relatively compact hadrons. In this case, a contribution
from the coulomb part of the Cornell confinement potential,
$V(r)=-\frac43\frac{\alpha_s}{r} + \sigma r$, is not small,
effectively "decreasing" the tension $\sigma$ at smaller distances
and, hence, masses of the ground $S$-wave states. In the case of
$\sigma$-meson, one can imagine the following situation: This
state represents a tetraquark but the admixture of additional
$q\bar{q}$-pair is small and gives a small direct contribution to
the mass. For this reason we may use the large-$N_c$ limit as a
first approximation. However, due to the extra $q\bar{q}$-pair,
the $\sigma$-meson (originally a scalar $P$-wave state) can exist
as a $S$-wave state. Due to this phenomenon, on the one hand, the
decay of this state becomes OZI-superallowed, explaining thereby
its abnormally large width, on the other hand, its mass decreases
similarly to the masses of ground $S$-wave vector mesons.

In our scheme, the value of slope of linear radial scalar
trajectory is taken from the solution of vector and axial planar
sum rules. This solution differs from the solution of
Ref.~\cite{peris}. According to the assumptions of
Ref.~\cite{peris}, the slopes of vector and axial trajectories are
different, as a consequence, the residues are also different, and
the quark condensate represents a input parameter (together with
$f_\pi$ and the gluon condensate). As a result, one has a system
of 8 polynomial equations for 8 variables: $\Lambda^2_{V,A}$,
$M^2_{V,A}$, $F^2_{V,A}$, $M_\rho^2$, and $F_\rho^2$. This system,
however, cannot be solved since it consists of two independent
groups of equations --- 4 equations for the vector channel and 4
for the axial one. The first group contains 5 variables and the
second one contains 3 variables. An approximate solution was found
by fixing $M_\rho$ and playing with $F_\rho$ in some range. We
believe that our ansatz and solution are more compact and natural.

The $\sigma$-meson within the large-$N_c$ Regge approach was also
studied in Ref.~\cite{arriola}, where it was found that the given
state represents a usual meson (it survives in the large-$N_c$
limit) and its mass lies in the interval 450 -- 600~MeV. These
conclusions agree with our results. However, the analysis made in
Ref.~\cite{arriola} is completely different. First, the
interpolating operator for the scalar isoscalar states was the
energy-momentum tensor in QCD. The results and conclusions were
heavily based on an analysis of the corresponding OPE of its
correlation function and some gravitational formfactors. Second,
all such states were placed on a single radial Regge trajectory
with half the standard slope. The existence of this possibility is
interesting but we believe that the predominantly non-strange and
predominantly strange isosinglet scalar mesons should form two
separate trajectories with approximately standard slope, as was
advocated in Ref.~\cite{phen}. Third, in the case of the
energy-momentum tensor, the scalar correlator should contain
additional poles corresponding to the glueball
states~\cite{narison}. They should cause a distortion of the pole
positions corresponding to the quark-antiquark states. For our
choice of the scalar interpolating current, we expect a
suppression of the glueball admixture in the planar limit. From
the phenomenological side, there is only one scalar candidate with
a presumably rich gluonic content --- the resonance
$f_0(1500)$~\cite{pdg}. We do not describe this state (in
particular, it is excluded from Fig.~1). The agreement in
estimating the $\sigma$-meson mass between our analysis and
Ref.~\cite{arriola} may be due to the fact that a natural glueball
scale where the distortion is maximal lies about 1~GeV higher than
the $\sigma$-mass.

It is interesting to observe that the old dual models
incorporating the chiral symmetry predict the degeneracy of the
radial vector and $f_0$ trajectories~\cite{avw}. This might be not
far from the reality, see Fig.~1.

The lattice calculations of $m_\sigma$ are still inconclusive.
Simulations with the simplest scalar quark current $J=\bar{q}q$ by
SCALAR Collaboration yielded a mass of the lightest ordinary
scalar isoscalar meson close to
$m_\rho\simeq0.77$~GeV~\cite{kunihiro}. An old simulation by Detar
and Kogut arrived at lower values~\cite{DeTar}. The work of SCALAR
Collaboration has recently been continued and the conclusion was
that the $\sigma$-meson may be a molecular state~\cite{wakayama}.
This conclusion, however, cannot be regarded as a serious evidence
against our results. The main findings of SCALAR Collaboration
consisted in the observation of a strong significance of
disconnected diagrams in the scalar isoscalar channel. In
addition, as correctly noticed in the Introduction of
Ref.~\cite{wakayama}, "the quark masses used in the present work
are admittedly not small, and hence it may not be straightforward
to extract direct implications regarding the nature of the sigma".
Indeed, in the simulations the authors had $m_\rho/m_\pi = 1.5$
while in the real world $m_\rho/m_\pi = 5.5$. One of conclusions
of the given analysis stated that for the comprehensive
understanding of the isosinglet scalar mesons the interpolation
operators including two-quark states and others should be taken
into account~\cite{wakayama}.

Our analysis was based on the standard OPE and the use of the
simplest scalar quark current. It is known that the scalar
correlator has also the so-called "direct instantons" contribution
(see, e.g., discussions in Ref.~\cite{steele1}). This contribution
is not seen in the OPE because of exponential fall-off. In
principle, the given contribution might lead to some non-linear
corrections to our linear spectrum. Perhaps the exponentially
decreasing corrections to the string-like spectrum introduced
phenomenologically in Ref.~\cite{we} could have a instanton
origin. A clarification of this issue represents a interesting
problem deserving a separate study.

It would be interesting to extend our analysis to the sector with
hidden strangeness. The combined sum rules for vector and axial
states will have a different numerical solution because the dim4
condensate $m_s\langle s\bar{s}\rangle$ is not negligible,
moreover, an effective dim2 condensate $m_s^2$ emerges from the
quark loop. Also the isovector sector with inclusion of the scalar
mesons $a_0$ should be considered. A study of these problems is
left for future.

\section{Conclusions}

We have considered the QCD sum rules in the large-$N_c$ limit
assuming for the radial excitations a linear Regge spectrum with
universal slope for the isosinglet vector, axial and scalar
mesons. The choice of spectrum is motivated by hadron string
models and related approaches and also by the meson spectroscopy.
The considered ansatz allows to solve the arising sum rules with a
minimal number of inputs. Since the QCD sum rules do not describe
microscopically neither the generation of QCD mass scale nor
spontaneous chiral symmetry breaking, the minimal number is two
and they parametrize numerically the given two phenomena. In our
scheme, the corresponding inputs are the gluon condensate and the
pion decay constant. The numerical solution of arising equations
reproduces the physical mass of $\omega(782)$-meson and a
consistent value for the quark condensate. The excited spectrum of
vector and axial states looks reasonable as well.

The obtained values of the slope of radial trajectories and quark
condensate are then used for the analysis of scalar channel. We
arrived at the conclusion that, interpolating the scalar states by
the simplest quark bilinear current, a prediction of light scalar
resonance with mass about $500\pm100$~MeV can be quite natural. We
indicated on the reasons of absence of this pole in the QCD sum
rules considered in the past. The coupling of this light scalar
meson to an external source does not reveal any unusual features.
It looks tempting to identify the given scalar state with
$f_0(500)$ which is commonly interpreted as a highly unusual
particle~\cite{pelaez}. This identification would mean that at
least the value of mass of $f_0(500)$ is not unusual. We also
observed that the mass of the lightest scalar meson, although not
being a part of the scalar radial Regge trajectory, correlates
strongly with the mass parameters of that trajectory. Concerning
the usual angular Regge trajectories for the quark-antiquark
states, we proposed a corresponding angular trajectory with
$f_0(500)$ on the top.

In summary, there is a possibility that the $\sigma$-meson
represents a "turnskin" resonance showing features of ordinary and
non-ordinary hadrons simultaneously. This makes revealing its
genuine nature even more challenging.

\section*{Acknowledgments}

The work was supported by the RFBR grant 16-02-00348-a and by the
Saint Petersburg State University.

\end{document}